\documentstyle[preprint,aps]{revtex}
\begin{document}
\draft
\title{Coherent stochastic resonance in the case of two
absorbing boundaries}
\author{Asish K. Dhara and Tapan Mukhopadhyay}
\address{Variable Energy Cyclotron Centre, 1/AF Bidhan Nagar, Calcutta-700064,
India}
\tighten
\date{\today}
\maketitle
\begin{abstract}
The coherent stochastic resonance is observed and studied with multi-step
periodic signal in continuous medium having two absorbing boundaries. The
general features of this process are exihibited. The universal features at
the resonance point are demonstrated. The kinetic behaviors around the
resonance point are also presented.
\end{abstract}
\section{introduction}
There has been a large deal of interest in the understanding of mechanism of
interplay between random noise and a deterministic periodic signal after the
pioneering achievement of separation of large DNA molecules in gel medium
by the application of uniform and time-dependent periodic electric field
\cite{sch,car}. It has been found that with this technique large molecules in
the size range 2 to 400 kb exihibit size-dependent mobilities. Similar
ideas have also arisen in other types of chromatographic processes \cite{lin}.

The first passage time is a useful tool to investigate the diffusive transport
property in a medium. The theory of first passage time has been worked out
in great detail for both infinite medium and explicitly time-independent
diffusive processes\cite{wei,gar,van}. However, for explicitly time-dependent
processes and in finite medium an analytic closed form expressions are not
available. In this respect also this problem attracts much attentions to the
scientific communities.

The first analysis of this phenomena has been done for a random walk on a
lattice numerically, and for a diffusive process in a continuous medium with
periodic signal of small amplitude perturbatively\cite{fle}. Their results
indicate that the oscillating field can create a form of coherent motion
capable of reducing the first passage time by a significant amount. This
fact clearly implies that the mobility of a particle in a diffusive medium
can be increased by the application of proper oscillating field. This phenomena
thereafter is known in the literatures as coherent stochastic resonance(CSR).

In order to investigate the reason for this cooperative behavior of random
noise and deterministic periodic signal this problem has been formulated
in much simpler terms by approximating the sinusoidal periodic signal by
the telegraph signal\cite{mas} and concluded incorrectly that the system
exihibits CSR. Subsequently it has been shown\cite{por} that the telegraph
signal can not produce CSR. It is then argued \cite{por} that the low frequency
behavior could cause such cooperative behavior.

In this paper we approximate the sinusoidal signal by a multi-step periodic
signal(explained below) and obtain an expression for the mean first
passage time(MFPT). After giving the derivation of MFPT in Sec.II, the results
of the calculations are discussed in Sec.III. First we present the general
characteristics of CSR. The calculation clearly exhibits how resonance
appears in our multi-step approximation and fails to show in single-step
telegraph approximation of the periodic signal explaining the conjecture
of Porra\cite{por}. The general characteristics of the moments in our
calculation are also in agreement with the numerical simulation of random
walk model on a lattice\cite{fle}. The characteristic features of first
passage time density function (FPTDF) for this phenomena are also
presented in this subsection. In the next subsection we focus on the resonance
point and demonstrate some universal features associated with it.  Subsequent
subsection deals with the characteristic changes of the physical variables
as we cross, in particular, around the resonance point. This leads to a better
understanding of this cooperative behavior. Finally, few concluding remarks
have been added in Sec.IV.
\section{Derivation of the mean first passage time}
We consider diffusion in one dimension perturbed by a periodic force. The
motion of the particle is given by the Langevin equation
\begin{equation}
\label{eq.1}
\dot{X} = A sin\Omega t + \xi(t),
\end{equation}
where $X$ refers to the stochastic variable, $A$ and $\Omega$ are the
amplitude and frequency of the sinusoidal signal and $\xi(t)$ is a zero mean
Gaussian white noise of strength $D$ with auto-correlation function given
by
\begin{equation}
\label{eq.2}
<\xi(t)\xi(t')> = 2D\delta(t-t').
\end{equation}
The motion is confined between two absorbing boundaries at $x=0$ and $x=L$.
The Fokker Planck equation corresponding to Eq.(\ref{eq.1}) is
\begin{equation}
\label{eq.3}
\frac{\partial p(x,t)}{\partial t} = - A sin\Omega t\frac{\partial p(x,t)}
{\partial x} + D\frac{\partial^2 p(x,t)}{\partial x^2},
\end{equation}
with absorbing boundary conditions at $x=0$ and $x=L$;i.e.,$p(0,t)=p(L,t)=0$.
We now introduce the dimensionless variables
\begin{equation}
\label{eq.4}
\xi=(A/D)x , \theta = (A^2/D)t , \omega = \Omega/(A^2/D) ,
\end{equation}
to write Eq.(\ref{eq.3}) in terms of new variables:
\begin{equation}
\label{eq.5}
\frac{\partial p(\xi,\theta)}{\partial\theta} = - sin\omega\theta
\frac{\partial p(\xi,\theta)}{\partial\xi} +\frac{\partial^2p(\xi,\theta)}
{\partial \theta^2}.
\end{equation}
The boundary conditions are rewritten as $p(0,\theta)=p(\Lambda,\theta)=0$,
where $\Lambda = (A/D)L$. In the following we calculate all the physical
quantities in terms of these new variables and if required, one may translate
all the interpretations in terms of the usual variables by the transformation
equations Eq.\ref{eq.4}.

We next approximate the sinusoidal signal by multi-step periodic signal.
The construction is as follows. We divide the half cycle of the signal by
 $(2p+1)$ intervals so that each interval in the horizontal $\theta$-axis
is of size $(\bigtriangleup\theta/(2p+1))$ with $\omega\bigtriangleup\theta
=\pi$. We define $(2p+1)$numbers $s_{k}$ along the vertical $\xi$-axis as
\begin{mathletters}
\label{eq.6}
\begin{eqnarray}
s_{k} &=& \frac{[ sin\frac{k\pi}{2p+1} + sin\frac{(k-1)\pi}{2p+1} ]}{2}~
; k = 1,2,...,p
\\s_{p+1} &=& 1
\\s_{p+1+r} &=& s_{p+1-r}~ ; r =1,2,...,p .
\end{eqnarray}
\end{mathletters}
Each number $s_k$ is associated with the interval$\frac{(k-1)\bigtriangleup
\theta}{2p+1} <\theta \leq \frac{k\bigtriangleup\theta}{2p+1}$ with
$k = 1,2,...,(2p+1)$. The Eq.(\ref{eq.6}) clearly shows that
\begin{equation}
\label{eq.7}
0<s_{1}<s_{2}<...<s_{p}<s_{p+1}=1>s_{p+2}>s_{p+2}>...>s_{2p+1}>0~.
\end{equation}
Eq.(\ref{eq.7}) states that in order to reach the maximum value $(=1)$ of the
signal from the zero level we have to have $(p+1)$ step up and from the
maximum to the zero level we have $(p+1)$ step down. This is for the positive
half-cycle. For the negative half-cycle similar constructions have been done
with the replacement $s_{k}\rightarrow - s_{k}, \forall k$ and each number
$- s_{k}$ is associated with the interval $\bigtriangleup\theta [1+
\frac{k-1}{2p+1}]<\theta\leq \bigtriangleup \theta [1+ \frac {k}{2p+1}]$
with $k=1,2,...,(2p+1)$. This aproximation for the full one cycle of the
sinusoidal signal (as shown in Fig.\ref{fig.1})is then repeated for the next
successive cycles. The construction clearly shows that we get back the usual
telegraph signal with $p=0$.

One may however note that the $\omega$ which we have defined for this
approximated signal is not the same as that of sinusoidal signal because
the Fourier transform of sinusoidal signal would give only one frequency
while this approximated signal in the Fourier space corresponds to many
sinusoidal frequencies specially because of its sharp discontinuities.
Yet we urge this approximation because in each interval the equation become
time-independent.

In the future development we associate the index $n$ for the positive
half-cycle and index $m$ for the negative. Index $i$ will refer the cycle
number. Since the Fokker-Planck equation (Eq.(\ref{eq.5})) in each interval
will be that for a constant bias, we can express the conditional probability
density function $p(\xi,\theta \mid \xi',\theta')$ in terms of complete
orthonormal set of eigenfunctions $u_{n}(\xi)$ satisfying the boundary
conditions $u_{n}(0) = u_{n}(\Lambda) =0$.
\begin{equation}
\label{eq.8}
p(\xi,\theta\mid\xi',\theta') = \sum_{n} u_{n}^+(\xi) u_{n}^-(\xi')
exp[-\lambda_{n}(\theta-\theta')]~,
\end{equation}
where
\begin{mathletters}
\label{eq.9}
\begin{eqnarray}
u_{n}^\pm(\xi) &=& (2/\Lambda)^\frac{1}{2}exp(\pm s\xi/2)sin\frac{
n\pi\xi}{\Lambda}~  ,\\
\lambda_{n} &=&\frac{n^2\pi^2}{\Lambda^2} +\frac{s^2}{4}~ .
\end{eqnarray}
\end{mathletters}
with $s$ as the corresponding value of $s_{k}$ in the appropriate interval
where the conditional probability is being decomposed. The conditional
probability density function in any interval, say $l$, can then be calculated
from the previous history by convoluting it in each previous intervals:
\begin{equation}
\label{eq.10}
p(\xi_{l},\theta_{l}\mid \xi_{1},\theta_{1}) =
\int...\int d\xi_{l-1}d\xi_{l-2}...d\xi_{2}\prod_{j=2}^{l}
p(\xi_{j},\theta_{j}\mid \xi_{j-1},\theta_{j-1})~.
\end{equation}
For the negative half-cycle the calculation of probability density function
is similar except that we have to replace the index $n$ by $m$ and the
probability density function is decomposed as
\begin{equation}
\label{eq.11}
p(\xi,\theta\mid\xi',\theta') = \sum_{m} u_{m}^-(\xi) u_{m}^+(\xi')
exp[-\lambda_{m}(\theta-\theta')]~,
\end{equation}
where the expressions for $u_{m}^\pm(\xi)$ and $\lambda_{m}$ are same as
in Eqs.(\ref{eq.9}).

The survival probability at time $\theta$ when the particle is known to
start from $\xi = \xi_0$ at $\theta = 0$ is defined as
\begin{equation}
\label{eq.12}
S(\theta\mid\xi_{0}) =\int_{0}^{\Lambda} d\xi p(\xi,\theta \mid \xi_{0},0)~.
\end{equation}

The first passage time density function (FPTDF) $g(\theta)$ is defined as
\begin{equation}
\label{eq.13}
g(\theta\mid\xi_{0}) = -\frac{dS(\theta\mid\xi_{0})}{d\theta}~.
\end{equation}
Physically, $g(\theta)d\theta$ gives the probability that the particle
arrives at any one of the boundaries in the time interval $\theta$ and
$\theta + d\theta$. From this density function one can calculate various moments:
\begin{equation}
\label{eq.14}
<\theta^{j}> = \int_{0}^\infty d\theta \theta ^{j} g(\theta)~.
\end{equation}
From Eq.(\ref{eq.14}) one can easily calculate mean first passage time(MFPT)
$<\theta>$ and the variance $\sigma^2 = <\theta^2>-<\theta>^2$ of the density
function $g(\theta)$.

It is then quite straight-forward to calculate the survival probability
at any interval of any cycle. We will write down the final formulae:

\label{eq.15}
\begin{mathletters}
\begin{eqnarray}
S_{+}(\theta\mid\xi_{0})=&& C^+_{n_{(2p+1)(i-1)+1}}
\times exp[-\lambda_{n_{(2p+1)(i-1)+1}}(\theta-2(i-1)\bigtriangleup\theta)]
\times F_{i-1}(u^-_{n_{(2p+1)(i-1)+1}})\nonumber\\
&&;2(i-1)\bigtriangleup\theta
<\theta\leq[2(i-1)+\frac{1}{2p+1}]\bigtriangleup\theta~,
\end{eqnarray}

\begin{eqnarray}
S_{+}(\theta\mid\xi_{0})=&& C^+_{n_{(2p+1)(i-1)+(k+1)}}
\times exp[-\lambda_{n_{(2p+1)(i-1)+(k+1)}}(\theta-2(i-1)\bigtriangleup\theta)]\times\nonumber\\
&&\prod_{j=1}^{k}\{<u^-_{n_{(2p+1)(i-1)+(j+1)}}\mid u^+_{n_{(2p+1)(i-1)+j}}>\}\nonumber\\
&&\times exp[\frac{\bigtriangleup\theta}{2p+1}(k\lambda_{n_{(2p+1)(i-1)+(k+1)}}
-\sum_{j=0}^{k-1}\lambda_{n_{(2p+1)(i-1)+(j+1)}})]\nonumber\\
&&\times F_{i-1}(u^-_{n_{(2p+1)(i-1)+1}})\nonumber\\
&&;[2(i-1)+\frac{k}{2p+1}]\bigtriangleup\theta
<\theta\leq[2(i-1)+\frac{(k+1)}{2p+1}]\bigtriangleup\theta\nonumber\\
&&;k=1,2,...,(2p-1)~,
\end{eqnarray}

\begin{eqnarray}
S_{+}(\theta\mid\xi_{0})=&& C^+_{n_{(2p+1)i}}
\times exp[-\lambda_{n_{(2p+1)i}}(\theta-(2i-1)\bigtriangleup\theta)]\nonumber\\
&&\times A^+(u^-_{n_{(2p+1)i}}, u^+_{n_{(2p+1)(i-1)+1}})
\times F_{i-1}(u^-_{n_{(2p+1)(i-1)+1}})\nonumber\\
&&;[2(i-1)+\frac{2p}{2p+1}]\bigtriangleup\theta
<\theta\leq(2i-1)\bigtriangleup\theta~,
\end{eqnarray}

\begin{eqnarray}
S_{-}(\theta\mid\xi_{0})=&& C^-_{m_{(2p+1)(i-1)+1}}
\times exp[-\lambda_{m_{(2p+1)(i-1)+1}}(\theta-(2i-1)\bigtriangleup\theta)]\nonumber\\
&&\times <u^+_{m_{(2p+1)(i-1)+1}}\mid u^+_{n_{(2p+1)i}}>\nonumber\\
&&\times A^+(u^-_{n_{(2p+1)i}}, u^+_{n_{(2p+1)(i-1)+1}})
\times F_{i-1}(u^-_{n_{(2p+1)(i-1)+1}})\nonumber\\
&&;(2i-1)\bigtriangleup\theta
<\theta\leq[(2i-1)+\frac{1}{2p+1}]\bigtriangleup\theta~,
\end{eqnarray}

\begin{eqnarray}
S_{-}(\theta\mid\xi_{0})=&& C^-_{m_{(2p+1)(i-1)+(k+1)}}
\times exp[-\lambda_{m_{(2p+1)(i-1)+(k+1)}}(\theta-(2i-1)\bigtriangleup\theta)]\times\nonumber\\
&&\prod_{j=1}^{k}\{<u^+_{m_{(2p+1)(i-1)+(j+1)}}\mid u^-_{m_{(2p+1)(i-1)+j}}>\}\nonumber\\
&&\times exp[\frac{\bigtriangleup\theta}{2p+1}(k\lambda_{m_{(2p+1)(i-1)+(k+1)}}
-\sum_{j=0}^{k-1}\lambda_{m_{(2p+1)(i-1)+(j+1)}})]\nonumber\\
&&\times <u^+_{m_{(2p+1)(i-1)+1}}\mid u^+_{n_{(2p+1)i}}>\nonumber\\
&&\times A^+(u^-_{n_{(2p+1)i}}, u^+_{n_{(2p+1)(i-1)+1}})
\times F_{i-1}(u^-_{n_{(2p+1)(i-1)+1}})\nonumber\\
&&;[(2i-1)+\frac{k}{2p+1}]\bigtriangleup\theta
<\theta\leq[(2i-1)+\frac{(k+1)}{2p+1}]\bigtriangleup\theta\nonumber\\
&&;k=1,2,...,(2p-1)~,
\end{eqnarray}

\begin{eqnarray}
S_{-}(\theta\mid\xi_{0})=&& C^-_{m_{(2p+1)i}}
\times exp[-\lambda_{m_{(2p+1)i}}(\theta-2i\bigtriangleup\theta)]\nonumber\\
&&\times A^-(u^+_{m_{(2p+1)i}}, u^-_{m_{(2p+1)(i-1)+1}})
\times <u^+_{m_{(2p+1)(i-1)+1}}\mid u^+_{n_{(2p+1)i}}>\nonumber\\
&&\times A^+(u^-_{n_{(2p+1)i}}, u^+_{n_{(2p+1)(i-1)+1}})
\times F_{i-1}(u^-_{n_{(2p+1)(i-1)+1}})\nonumber\\
&&;[(2i-1)+\frac{2p}{2p+1}]\bigtriangleup\theta
<\theta\leq 2i\bigtriangleup\theta~,
\end{eqnarray}
\end{mathletters}
where
\begin{mathletters}
\begin{eqnarray}
\label{eq.16}
C_{n}^+ =&&\int_{0}^{\Lambda} d\xi u_{n}^+(\xi)~,
\end{eqnarray}

\begin{eqnarray}
C_{m}^- =&&\int_{0}^{\Lambda} d\xi u_{m}^-(\xi)~,
\end{eqnarray}

\begin{eqnarray}
A^+(u^-_{n_{(2p+1)i}}, u^+_{n_{(2p+1)(i-1)+1}})=&&exp[-(\frac{\bigtriangleup\theta}{2p+1})\lambda_{n_{(2p+1)(i-1)+1}}]\nonumber\\
&&\prod_{j=1}^{2p}\{<u^-_{n_{(2p+1)(i-1)+(j+1)}}\mid u^+_{n_{(2p+1)(i-1)+j}}>\nonumber\\
&&\times exp[-(\frac{\bigtriangleup\theta}{2p+1})\lambda_{n_{(2p+1)(i-1)+(j+1)}}]\}~,
\end{eqnarray}

\begin{eqnarray}
A^-(u^+_{m_{(2p+1)i}}, u^-_{m_{(2p+1)(i-1)+1}})=&&exp[-(\frac{\bigtriangleup\theta}{2p+1})\lambda_{m_{(2p+1)(i-1)+1}}]\nonumber\\
&&\prod_{j=1}^{2p}\{<u^+_{m_{(2p+1)(i-1)+(j+1)}}\mid u^-_{m_{(2p+1)(i-1)+j}}>\nonumber\\
&&\times exp[-(\frac{\bigtriangleup\theta}{2p+1})\lambda_{m_{(2p+1)(i-1)+(j+1)}}]\}
\end{eqnarray}
\end{mathletters}
and the function $F_{i}$ is generated through the recursion relation:

\begin{eqnarray}
\label{eq.17}
F_{i}(u^-_{n_{(2p+1)i+1}})=&&<u^-_{n_{(2p+1)i+1}}\mid u^-_{m_{(2p+1)i}}>
\times A^-(u^+_{m_{(2p+1)i}}, u^-_{m_{(2p+1)(i-1)+1}})\nonumber\\
&&\times <u^+_{m_{(2p+1)(i-1)+1}}\mid u^+_{n_{(2p+1)i}}>
\times A^+(u^-_{n_{(2p+1)i}}, u^+_{n_{(2p+1)(i-1)+1}})\nonumber\\
&&\times F_{i-1}(u^-_{n_{(2p+1)(i-1)+1}})~,
\end{eqnarray}

with $F_{0}(u^-_{n_{1}}) = u^-_{n_{1}}(\xi_0)$. The angular bracket in any equation implies dot product of the corresponding
functions,for e.g.,
\begin{equation}
\label{eq.18}
<u^+\mid u^-> = \int_{0}^{\Lambda} d\xi u^+(\xi)u^-(\xi)~.
\end{equation}
The cycle variable $i$ runs over positive integers;i.e.,$i=1,2,3,...$. The
positive and negative symbols of the survival probabilities indicate its
value over positive and negative part of the cycles respectively. In all
these expressions, viz., Eqs.(15)-(\ref{eq.17}), any subscript either
$n$ or $m$ or both whereever they appear more than once the summation over
them are implied. The effect of history is explicit in the expressions for
survival probabilities. Once the survival probability $S(\theta\mid\xi_0)$
is obtained from these formluae, the FPTDF,MFPT and the corresponding variance
are obtained by employing Eqs.(\ref{eq.13})-(\ref{eq.14}).
Evaluation of MFPT and other relevant quantities requires sum of infinite
series which must be truncated in order to obtain a final result. Convergence 
of MFPT is ensured by gradually increasing the number of terms (i.e.,number
of eigenvalues) for the calculation. The process is truncated when MFPT
does not change upto two decimal point of accuracy with the change of number
of terms.
\section{results and discussions}
The survival probability, mean first passage time(MFPT), corresponding
variances and first passage time density functions(FPTDF) are calculated
using the derived formulae for this process. The results are summarised below.
\subsection{General features of CSR}
The MFPT is calculated for single-step telegraph signal $(p=0)$ with
$\xi_0=\Lambda/2$. Most of the calculations are done with this specific
value of $\xi_0$. The variation of the results with variation of $\xi_0$
is also demonstrated [see the text below]. No
nonmonotonous behavior is observed in MFPT as we vary the frequency $\omega$.
This is in complete agreement with Porra's observation\cite{por}. The calculation
is done for the length $\Lambda = 20$ and the result is shown in the curve
$a$ of Fig.2.However, when we take $p=1$, i.e., when the sinusodal
signal is approximated by two-step periodic signal, the calculation of
MFPT for the same length shows clearly the nonmonotonous behavior. This is
shown in curve $b$ of the same figure. This result clearly demonstrates that
mere flipping of the bias (signal) direction periodically would not produce
the coherent motion. As the rate of flipping increases it merely prevents the
particle more to reach the boundaries and therefore MFPT increases
monotonically. It may be noted that when the flipping rate is very high, the
effect of signal is almost nil and the transport is effectively diffusive in
nature. This is of course true in any type of periodic signal. Therefore, for
any type of approximation of the sinusoidal signal or for any value of $p$,
this feature would show up. In particular, for $p=1$, we observe from curve $b$
of fig.2 that MFPT asymptotically reaches the diffusive limit $\Lambda^2/8$
(= 50 in this case). The usual
telegraph signal offers a constant bias of maximum magnitude for the larger
time than for a two-step approximation. Hence the particle always has a larger
probability of reaching the boundary in short time for $p=0$ case than for
$p>0$ case. Hence MFPT for $p=0$ and for any $\omega$ is always less than for
$p>0$ case. This is observed in Fig.2.

The application of any bias always reduces the MFPT than for the non-biased
diffusion. In CSR we always have a competition between diffusion and
oscillatory effect of the bias. For very large frequency as the bias effect
becomes ineffective MFPT would essentially be guided by diffusive process.
For zero frequency of the multi-step periodic signal the MFPT can be
analytically evaluated. When it starts from the mid-point of the medium it
expresses as $<\theta(\omega=0,\xi_0=\Lambda/2)>
=0.5(\Lambda/s_1)tanh(s_1\Lambda/4)$.
When frequency is very small, the process is predominantly diffusion with
constant value $s_1=0.5 sin(\frac{\pi}{2p+1})$ effective for
$0<\theta\le\frac{\pi}{\omega(2p+1)}$.
However, as frequency
increases slowly the probability of having increased bias value $s_2$  
(=1 for $p=1$) before it reaches the boundary increases. This bias force reduces the
survival probability and also
MFPT. Hence one would expect a minimum to MFPT. On the otherhand, for usual
telegraph signal ($p=0$ case), for very low frequency, from the very beginning
bias force affects the particle with its maximum strength. When the frequency
is very low, this constant bias diffusion continues for a longer time and there
is no change-over of the magnitude of the bias as in the case of $p=1$.
After having a flip, the particle again suffers a constant bias diffusion
in the direction opposite to the previous one.
As frequency increases slowly, this picture remains unchanged until a stage
reaches for which the flipping effect becomes dominant during the particle's
survivality inside the medium and MFPT increases. This is observed in Fig.2.

Next we continue all our calculation with $p=2$ or, with three-step telegraph
signal. Calculation reveals that the value of MFPT does not change much from
that with $p=1$. On the otherhand, $p=2$ signal approximates better than $p=1$
signal. We restrict our calculation with $p=2$ approximation of the periodic
signal.

Typical survival probability and the corresponding decay rate defined as
$\rho (\theta) = -\frac{dS(\theta)}{d\theta}/S(\theta)$ are plotted as
function of $\theta$ for $\Lambda = 20$ and $\omega=0.1$ in Fig.3.
The plot shows that the survival probability [plot a] goes through plateau where the
change of survival probability is comparatively less. The decay rate $\rho
(\theta)$ [plot b] correspondingly shows a minimum at these points. This is a
characteristic feature for CSR. This feature is in agreement with the numerical
simulation of the process as a random walk on a lattice\cite{fle}.

Next we calculate the MFPT $<\theta>$ and the variance $\sigma^2$ as a function
of frequency $\omega$ for different lengths$(\Lambda = 10,20,30,40,50)$.
These are presented in Fig.4 and Fig.5 respectively. Both the cumulants go
through a minimum as frequency rises from very low value for each length
$\Lambda$. This feature is also in agreement with the lattice simulation work
\cite{fle}. It is observed that the minimum for both the moments ocuur at the
same frequency for each length. The value of MFPT $<\theta>$ increases with the
length at all frequencies. This is understandable because as length increases
on an average the particle will spent more time in the medium before reaching
the boundaries. It is also observed that the frequency at which the minimum
occurs shift towards low frequency as the length increases. It implies that
maximum cooperation between the deterministic signal and random noise occurs
at lower frequencies as the length increases. For low resonant frequency the
particle is affected by the bias in a particular direction for a longer
period of time before it suffers a change in the direction of bias, thus
more probability to cover a large distance towards the boundary and at this
resonant frequency the probability for reaching the boundary in a short time
is maximum because if one increases the frequency more than the resonant
frequency at that length, the flipping rate dominates and average time taken
by the particle would be more.

Fig.5 demonstrates the lowering of the dispersion at resonant frequencies
confirming that the cooperation is maximum at these frequencies. Dispersion is
more for higher lengths and as seen from the figure the dispersion merges to a
specific value at very low frequency at various lengths. 

All the previous calculations are done when the particle starts initially
from the mid point of the medium, i.e., $\xi_{0}$ in Eqs.(15) is taken as
$\Lambda/2$. At the length $\Lambda=20$ the resonant frequency is found to be
0.1. The calculations are done one at resonant frequency and other two at the off-
resonant frequencies ($\omega=0.5$ and $\omega=0.0$) 
when the particle starts from $\xi_{0}=\beta\Lambda$ where
$\beta$ lies between 0 and 1.For zero frequency the MFPT can be analytically
obtained. Its expression reads as $<\theta(\omega=0,\beta\Lambda)>=
(\Lambda/s_1)[-\beta+(1-exp[-s_1\beta\Lambda])/(1-exp[-s_1\Lambda])]$. 
The curves are shown in Fig.6. It is evident
that the value of $<\theta>$ is less for resonant frequency (curve $a$) than
for its value for off-resonant frequencies(curves $b$ and $c$). As frequency increases,
the maximum value of $<\theta>$ occurs at lower values of $\beta$ or, when the
particle starts from the left of the interval. It is known that for pure
diffusion the location of maximum $<\theta>$ would occur for $\beta=0.5$.
Our signal starts with positive half-cycle and therefore the survival time
of the particle would be more if the particle starts from the left of the
interval. Of course there would be some limit, because if it starts too much
near the left end then diffusion towards the left boundary dominates and
average time would be less. On the otherhand, if it starts from right half of
the medium, the initial surge of the signal helps the particle to reach the
boundary more quickly. Hence average time of duration decreases. This fact
is also in conformity with lattice simulation work\cite{fle}, although
most of the simulations in \cite{fle} were obtained for an uniformly
distributed initial condition.

We next calculate the FPTDF $g(\theta)$ for various frequencies for $\Lambda
=20$ and plot the curves in Fig.7. The resonant frequency for this length is
found to be 0.1. Before the resonant frequency is reached, $g(\theta)$ has got
two distinct peaks [Fig.7a] and at resonance two peaks merge to a single large
peak. After the resonance many smaller peaks in $g(\theta)$ gradually emerge
as frequency increases more than the resonant frequency [Fig.7b]. This is
a general characteristic of CSR. The height $h$ and the position $\theta_{p}$
of the first peak as a function of $\omega$ are plotted in curve $a$ of Fig.8.
The figure shows that height of the first peak goes through a maximum as we
increase the frequency while the position of that peak remains practically
constant. The height reaches the maximum near the resonant frequency demonstrating that
the probability of reaching the boundary is in short time is maximum
near the resonant frequency. It is a kind of reflection of having $<\theta>$
minimum at that frequency. Therefore it is a general characteristic of CSR.
The height and position of the second peak before the resonance are drawn
as curve $b$ in Fig.8. At resonance the two peaks merge and we have only one
peak. Just after the resonance another peak starts developing and height
increases as frequency increases further. The position and height of the second
peak after resonance are plotted in curve $c$ of the same figure. The merging
and the reappearance of the second peak is also observed as a brake or
discontinuity of the dashed line in this figure.
\subsection{Universal features at resonance}
In this subsection we concentrate on the behaviour of the system at the
resonance point. We have already discussed some general characteristics of CSR
in the previous subsection. We find that for each length, $\Lambda$, a
corresponding frequency $\omega^*$ exists for which $<\theta>$ and $\sigma^2$
become minimum implying that the maximum cooperation between the deterministic
periodic signal and random noise of the environment is taking place in
helping the particle to reach the boundaries. One therefore would naturally
inquire about the relation of $\omega^*$ with $\Lambda$. The curve of
$\omega^*$ as a function of $\Lambda$ is plotted in Fig.9. In the range of
$\Lambda$ we studied this curve is very well fitted with the formula
\begin{equation}
\label{eq.19}
\omega^* = 2/\Lambda~.
\end{equation}

The values of MFPT at resonance $<\theta(\omega^*)>$ is plotted against the
length $\Lambda$ in Fig.10 and within the range of $\Lambda$ we consider
the relation between them is fitted to
\begin{equation}
\label{eq.20}
<\theta(\omega^*)> = 0.82 \Lambda -0.14~.
\end{equation}
Of course, there will be deviation from this linear behaviour as $\Lambda$
decreases further because $<\theta>$ can not become negative and for
$\Lambda=0$(corresponding to $L=0$), $<\theta>$ should be zero.

Similarly the variance $\sigma^2(\omega^*)$ is plotted as a function of
$\Lambda$ in Fig.11 and within the range of $\Lambda$ we consider this curve
is fitted to
\begin{equation}
\label{eq.21}
\Lambda =a[\sigma^2(\omega^*)]^2 + b~,
\end{equation}
with $a=.004,b=9.29\pm 0.82$.

We have already seen that at the resonance frequency we have one very dominant
peak of FPTDF, $g(\theta)$[Fig.7b]. Since it is a general feature, for each
length $\Lambda$ we should get such behaviour. We further observe that
$\omega^*$ varies inversely with $\Lambda$ (Eq.(19)). With this fact in our
mind when we plot $g(\theta)/\omega^*$ as a function of $[\omega^*(\Lambda)\theta]$,
we find that curves for all lengths superpose over each other [Fig.12] and the pattern
of $g(\theta)/\omega^*$ for different $\Lambda$ or $\omega^*$ is very similar, i.e.,
at particular values of $[\omega^*\theta]$, all curves show their maxima,
minima, and change in the behavioral patterns of the curves occur exactly
at the same places of$[\omega^*\theta]$. Similar characteristics are also
observed in the curves of decay rate $\rho$ for different frequencies. For
illustration we plot $\rho$ as a function of $[\omega^*\theta]$ for three
different lengths($\Lambda=20$ ; curve $a$, $\Lambda=28.57$; curve $b$,
$\Lambda=15.38$ ; curve $c$), and present in Fig.13. Therefore it shows that this
feature is universal and $[\omega^*\theta]$ or the cycle number is the correct
variable to describe the resonance behaviour. We may further note that such
scaling of FPTDF would not be possible for any frequency other than the
resonant frequencies because any frequency which is not the resonant frequency
for one length may turn out to be the resonant frequency for some other 
length and the features of FPTDF are different for resonant and off-resonant
frequencies as has been observed from Fig.7a - Fig.7b. The major dominant peaks of
FPTDF $g(\theta)/\omega^*$ for different lengths$(\Lambda = 20,28.57,35,40,44.44,50)$
are drawn as a function of $[\omega^*\theta]$ in Fig.12. There are no overlap
of these curves which can be seen as we increase the resolution. The lowermost
curve is for $\Lambda = 20$ and as length increases the upper curves are
generated. The peaks for all the curves occur nearly at a quarter of a cycle.

The peak height $h_{p}$ and full width half maximum(FWHM) for each curve is
plotted as a function of resonant frequency. The plot is given in Fig.14. The
plot shows that except for very low frequency they behave linearly with
the resonant frequency.
\subsection{Behaviour around the resonant point}
We have already seen that the cooperation between the deterministic signal
and the random noise is maximum at the resonance point where MFPT $<\theta>$
variance $\sigma^2$ take minimum values and corresponding FPTDF $g(\theta)$
shows a major dominant peak. What would happen when we change the frequency
slightly above and below the resonant frequency? To investigate the matter
we choose a particular length of the medium, $\Lambda = 20$. The resonance
frequency for such length, $\omega^*=0.1$. For this particular length we
take two off-resonant frequencies $\omega = 0.07$ and $\omega = 0.13$. The
curves for survival probabilities as a function of time $\theta$ are plotted
in Fig.15, where the curves $a, b, c$ are for frequencies $\omega =0.1,.07,.13$
respectively. The calculation of survival probability is terminated when it takes value
$1\times 10^{-3}$ which corresponds to zero in our calculation. The curves
clearly show that as frequency increases the survivality of the particle
prolongs. This is quite understandable because more oscillations prevent the
particle to reach the boundary, i.e., for higher frequency we expect $<\theta>$
more. The oscillatory effect is more pronounced when time is large. For large
time we always expect the value of $S(\theta)$ more for higher frequency.
This is clearly observed in Fig.15. But for frequency lower than the resonant
frequency MFPT $<\theta>$ is again more. As MFPT is the integral of the survival
probability over time, we expect a change in the behaviour of $S(\theta)$
for lower time regime. This is shown explicitly in Fig.16. In this figure
we find that the survival probability is more for low frequency, $\omega = 0.07$
(curve $b$) than for resonant frequency, $\omega^* = 0.1$(curve $a$) and
off-resonant frequency $\omega = 0.13$(curve $c$). Especially for curve $b$,
the value of $S$ is so much than that for curve $a$, so that area under the
curve $b$ is more than that for curve $a$. We see that near about $\theta=28$,
the solid curve crosses the dashed curve. There is only one point of crossing
throughout the entire time. We have already argued that after this crossing
point oscillatory effect of this bias dominates. It is then clear for low
time regime diffusion process competes over the oscillatory effect. Again,  for
very low $\theta$ the chance of having increased value of the bias
in the same direction is more for high frequency than for low frequency. Therefore
for low frequency the survivality is more than for the high frequency. The
curves in Fig.16 also demonstrate that.

We have already demonstrated how $<\theta(\omega^*)>$ varies with $\Lambda$
in Fig.10. The behaviour is linear with respect to the length of the medium.
It is of interest whether the behaviour is changed for off-resonant frequency.
For that we choose a frequency which is not the resonant frequency for the
length $\Lambda$ that we consider in our calculation. The MFPT $<\theta(\omega)>$
for that off-resonant frequency are calculated for different lengths and are
plotted as a function of $\Lambda$ in Fig.17. For the range of length we
consider the curve is fitted to
\begin{equation}
\label{eq.22}
\Lambda = a'<\theta(\omega_{off-res})>^2 + b'
\end{equation}
with $a' =.016$ and $b'= 8.68\pm.33$. We may note that this particular
off-resonant frequency would be a resonant frequency for some length, $\Lambda_0$,
governed by the Eq.(19). In our case this off-resonant frequency corresponds
to length $\Lambda_0 > 50$. The curve shows that when $\Lambda \ll \Lambda_0$,
the rate of change of MFPT with respect to $\Lambda$ is more and when $\Lambda$
approaches $\Lambda_0$ or, when this frequency tends to be the resonant frequency,
the rate is curbed. This could be a signature of approaching a coherent motion
from the non-cooperative behaviour.
\section{concluding remarks}
We consider a diffusive transport process perturbed by a periodic signal in
continuous one dimensional medium having two absorbing boundaries. No
perturbation of the signal amplitude is assumed in this formulation. We showed
explicitly that the cooperative behaviour between the deterministic periodic
signal and random noise leading to coherent motion occurs when the
time-dependent sinusoidal signal is approximated by a multistep periodic
signal and not with single-step telegraph signal.

Although we study the process with three-step periodic signal, the formulation
is quite general and applicable for any approximation with arbitrary number of steps.
This formulation can also be applied to any arbitrary continuous periodic
signal.

It is observed that for large time oscillation of the signal plays a dominant
role in the transport while in the low time regime frequency dependent bias
force (i.e.,the chance of having increased values of the bias in the same
direction is more for high frequency than for low frequency)
has the key factor. For very high frequency the bias effect is
practically absent and the motion is purely diffusive in nature. At the
resonance the maximum cooperation between the noise and the periodic signal
takes place.

An important characteristic that we observe is that at the resonance the
FPTDF for various lengths have similar behaviour as a function of cycle
number. There is only one dominant peak and the peak position occurs very near
to a quarter of a cycle. From Fig.12 we observe a slight deviation of the peak
positions but we believe that if the sinusoidal signal is approximated
by more than three-step periodic signal the position of all the peaks
will be the same.

There is also slight discrepancy in the position of the minimum of
$\sigma^2$ in comparison to the minima of $<\theta>$[Fig.4,Fig.5]. This may
be due to the fact that all calculations are made to an end when the
survival probability takes a value $1\times 10^{-3}$. We observe that if
we cut off the calculations for more lower values of survival probability
it does not affect MFPT but the variances are slightly affected. Also
if one approximates the sinusoidal signal better than three-step periodic
signal one could obtain the positions of the minima of variances at exactly
the same places as those with MFPT.

It is intersting to observe that the decay rate at the resonance [Fig.13]
after $\omega^* \theta =5\pi/4$ is clearly a periodic function of time.

\begin{figure}
\caption{Sinusoidal signal(dashed curve) and approximated three-step(p=2)
periodic signal(solid curve) for the full one cycle as a function of $\theta$.}
\label{fig.1}
\end{figure}

\begin{figure}
\caption{MFPT $<\theta>$ as a function of $\omega$ ;
(a)for p=0 ;the usual telegraph
signal (b) for p=1 ; the two-step periodic signal,
[$\Lambda=20,\xi_0=\Lambda/2$].}
\label{fig.2}
\end{figure}

\begin{figure}
\caption{(a) $-ln S(\theta)$ as a function of $\theta$ (dashed curve)
(b) The decay rate, $\rho$, as a function of $\theta$ (solid curve),
[$\Lambda=20, \xi_0=\Lambda/2, p=2, \omega=0.1$].}
\label{fig.3}
\end{figure}

\begin{figure}
\caption{MFPT $<\theta(\omega)>$ as a function of frequency $\omega$ ;
(a)$\Lambda=10$, (b)$\Lambda=20$, (c)$\Lambda=30$, (d)$\Lambda=40$, (e)$\Lambda=50$,
[$p=2,\xi_0=\Lambda/2]$.}
\label{fig.4}
\end{figure}

\begin{figure}
\caption{The variance $\sigma^2$ as a function of $\omega$ ;
(a)$\Lambda=10$, (b)$\Lambda=20$, (c)$\Lambda=30$, (d)$\Lambda=40$, (e)$\Lambda=50$,
[$p=2,\xi_0=\Lambda/2]$.}
\label{fig.5}
\end{figure}

\begin{figure}
\caption{MFPT $<\theta>$ as a function of $\beta$ for length $\Lambda=20$ ;
(a) for resonant frequncy $\omega^* = 0.1$
(b) for off-resonant frequncy $\omega = 0.5$
(c) for off-resonant frequncy $\omega = 0.0$,
[p=2]. }
\label{fig.6}
\end{figure}

\begin{figure}
\caption{(a) FPTDF $g(\theta)$ as a function of $\theta$ for $\Lambda=20$
before resonance for frequencies $\omega$ = .01,.02,.03,.07
(b) FPTDF $g(\theta)$ as a function of $\theta$ for $\Lambda=20$
on and after resonance for frequencies $\omega = 0.1$(resonant),
$\omega =.13, 0.2, 0.3 $ respectively,[$p=2,\xi_0=\Lambda/2]$.}
\label{fig.7}
\end{figure}

\begin{figure}
\caption{Height $h$ and the position of the peak $\theta_p$ as a function
of $\omega$ ;
(a) for the first peak(solid curve)
(b) for the second peak before resonance(dashed curve)
(c) for the second peak after resonance (dotted curve).}
\label{fig.8}
\end{figure}

\begin{figure}
\caption{Resonant frequency $\omega^*$ as a function of length $\Lambda$.}
\label{fig.9}
\end{figure}

\begin{figure}
\caption{MFPT at the resonant frequency $<\theta(\omega^*)>$ as a function
of length $\Lambda$. }
\label{fig.10}
\end{figure}

\begin{figure}
\caption{The variance at the resonant frequency $\sigma^2(\omega^*)$ as a
function of the length $\Lambda$.}
\label{fig.11}
\end{figure}

\begin{figure}
\caption{The dominant peaks of $g(\theta)/\omega^*$ at resonant frequencies
for different lengths ($\Lambda = 20, 28.57, 35,40, 44.44, 50$) as a function 
of $\omega^*\theta$. The lowermost curve is for $\Lambda = 20$ , and as length
increases gradually upper curves are generated,[$p=2,\xi_0=\Lambda/2]$.}
\label{fig.12}
\end{figure}

\begin{figure}
\caption{The decay rate $\rho$ for different resonant frequencies as a
function of $\omega^*\theta$;
(a) $\Lambda=20, \omega^* = 0.1$ (solid curve)
(b) $\Lambda=28.57, \omega^* = 0.07$ (dashed curve)
(c) $\Lambda=15.38, \omega^* = 0.13$ (dotted curve),[$p=2,\xi_0=\Lambda/2]$.}
\label{fig.13}
\end{figure}

\begin{figure}
\caption{The height $h_p$ and full width half maximum FWHM of the peaks
in FIG.12 are plotted as a function of their corresponding resonant
frequencies.}
\label{fig.14}
\end{figure}

\begin{figure}
\caption{$-lnS(\theta)$ as a function of time $\theta$ ;
(a) $\Lambda=20, \omega^* = 0.1$ (solid curve)
(b) $\Lambda=20, \omega^* = 0.07$ (dashed curve)
(c) $\Lambda=20, \omega^* = 0.13$ (dotted curve),[$p=2,\xi_0=\Lambda/2]$.}
\label{fig.15}
\end{figure}

\begin{figure}
\caption{$S(\theta)$ as a function of $\theta$ for the same curves as in
FIG.15.}
\label{fig.16}
\end{figure}

\begin{figure}
\caption{MFPT at off-resonant frequency $<\theta(\omega_{off-res.})>$ as a
function of $\Lambda$,[$p=2,\xi_0=\Lambda/2]$.}
\label{fig.17}
\end{figure}

\begin{references}
\bibitem{sch}D.C.Schwartz and C.R.Cantor, Cell {\bf37}, 67 (1984).
\bibitem{car}G.F.Carle, M.Frank, and M.V.Olson, Science {\bf232}, 65 (1986).
\bibitem{lin}I.J.Lin and L.Benguigi, Sep.Sci.Tech. {\bf20}, 359 (1985).
\bibitem{wei}G.H.Weiss, Adv.Chem.Phys. {\bf13}, 1 (1967).
\bibitem{gar}C.W.Gardiner, {\it A handbook of Stochastic Methods},
2nd ed.(Springer-Verlag, New York, 1985).
\bibitem{van}N.G.van Kampen, {\it Stochastic processes in physics and
chemistry} (North-Holland, Amsterdam, 1991).
\bibitem{fle}J.E.Fletcher, S.Havlin, and G.H.Weiss, J.Stat.Phys. {\bf51},
215 (1988).
\bibitem{mas}J.Masoliver, A.Robinson, and G.H.Weiss, Phys.Rev. E{\bf51},
4021 (1995).
\bibitem{por}J.M.Porra, Phys.Rev. E{\bf55}, 6533 (1997).
\end{references}
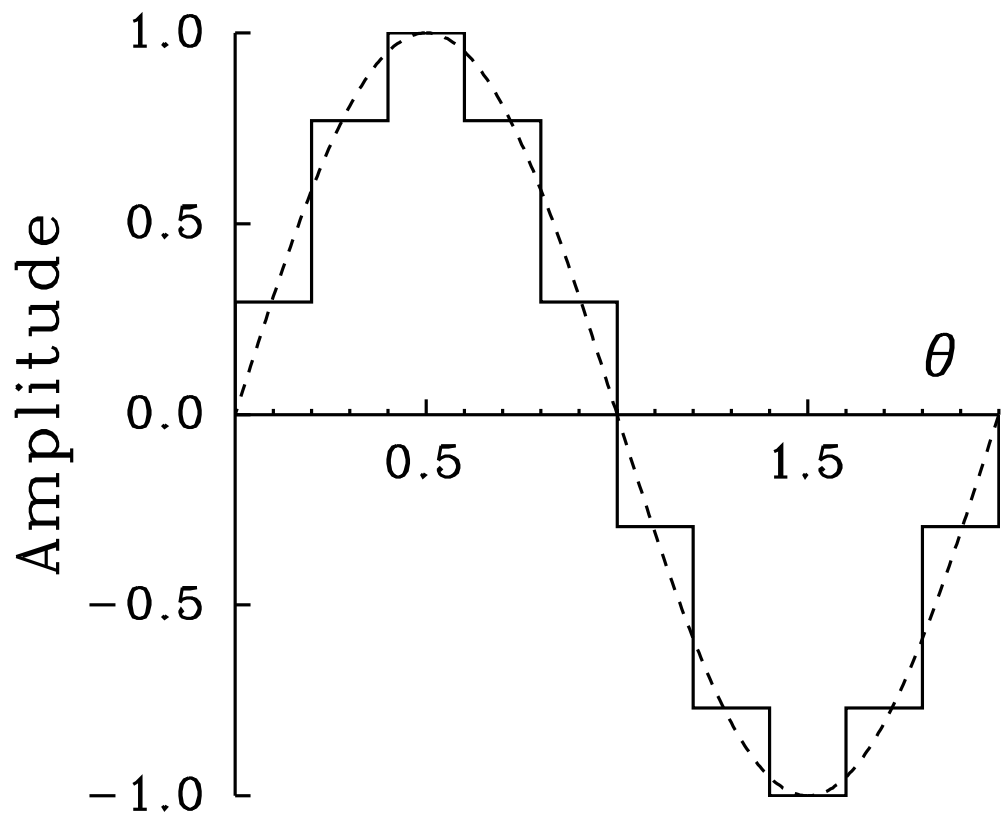
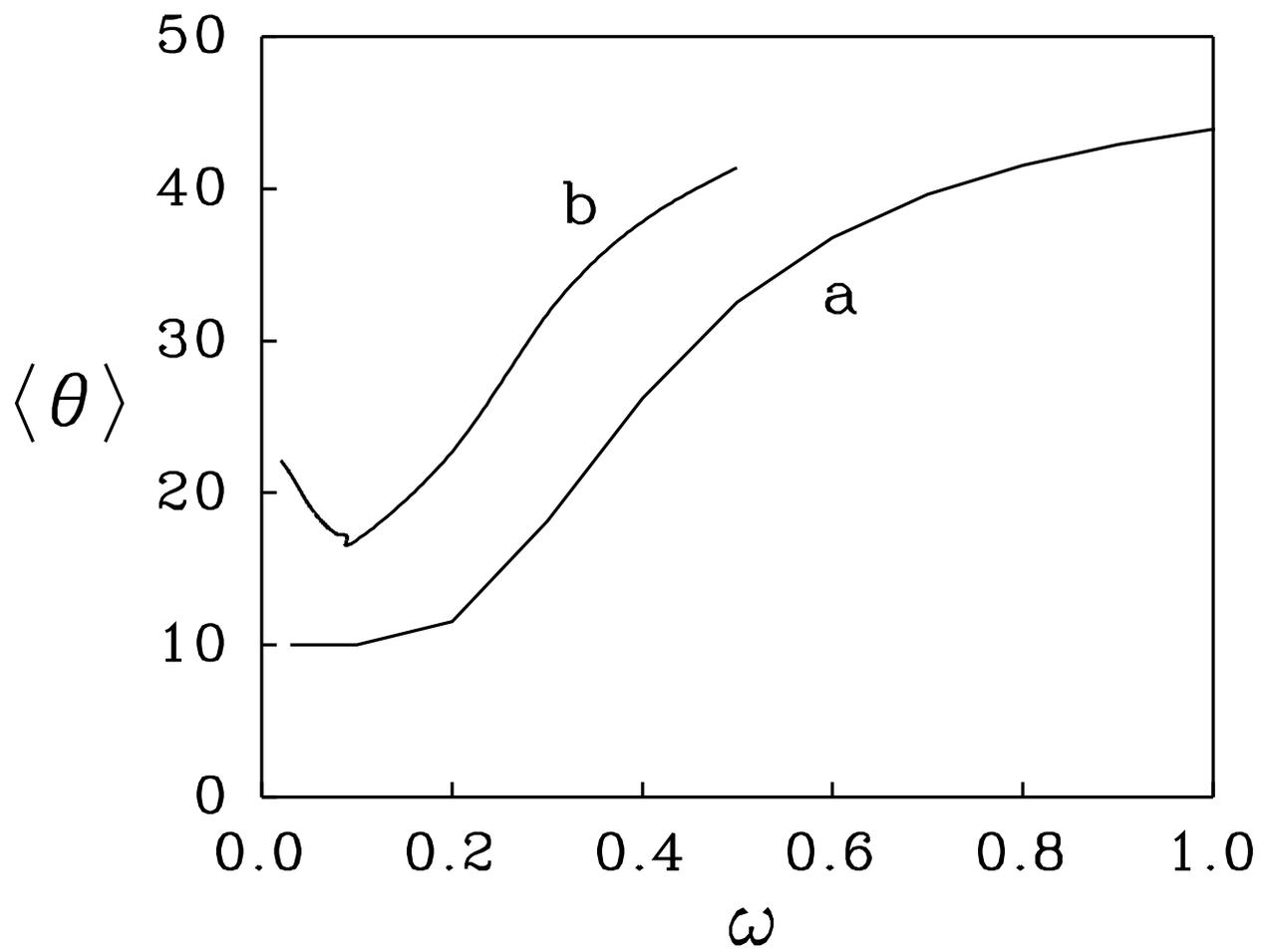
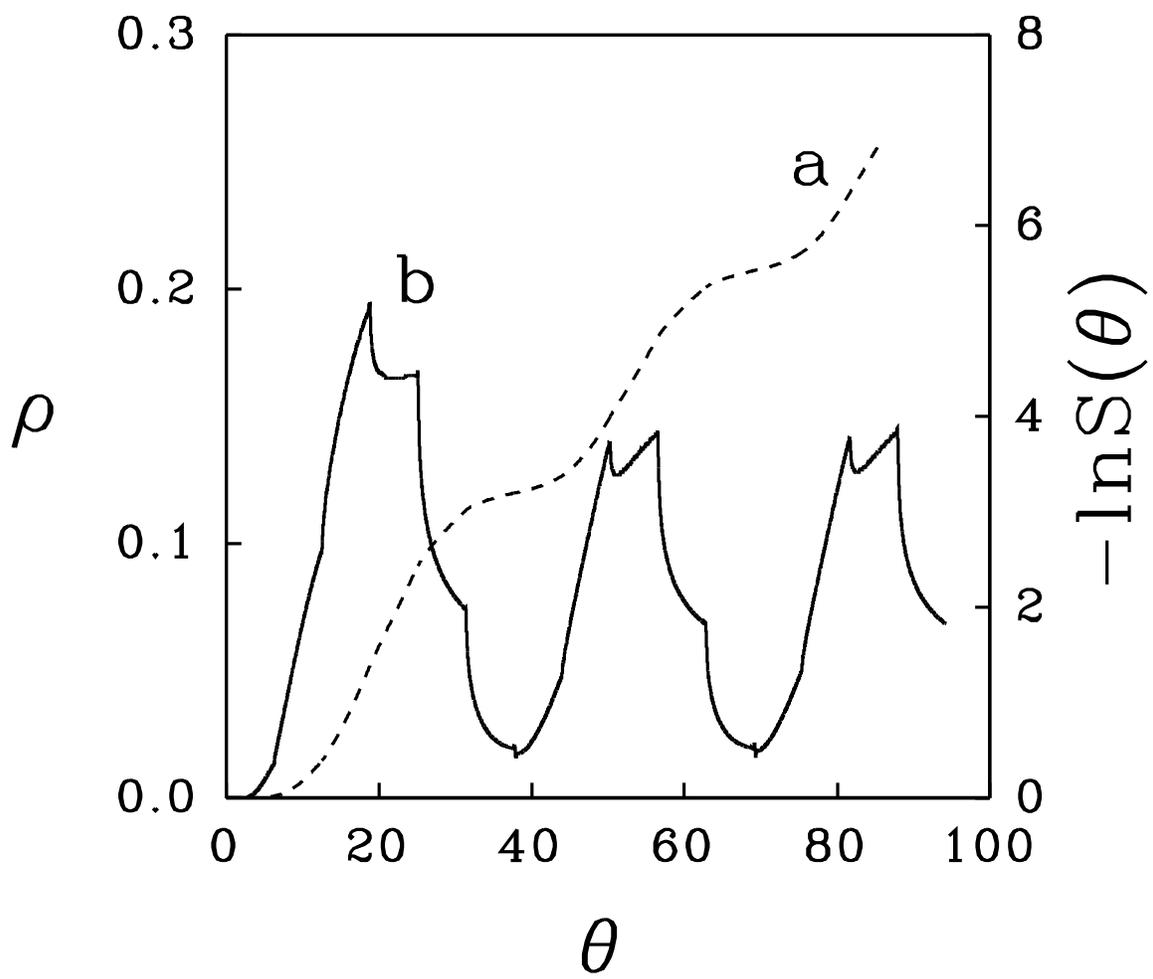
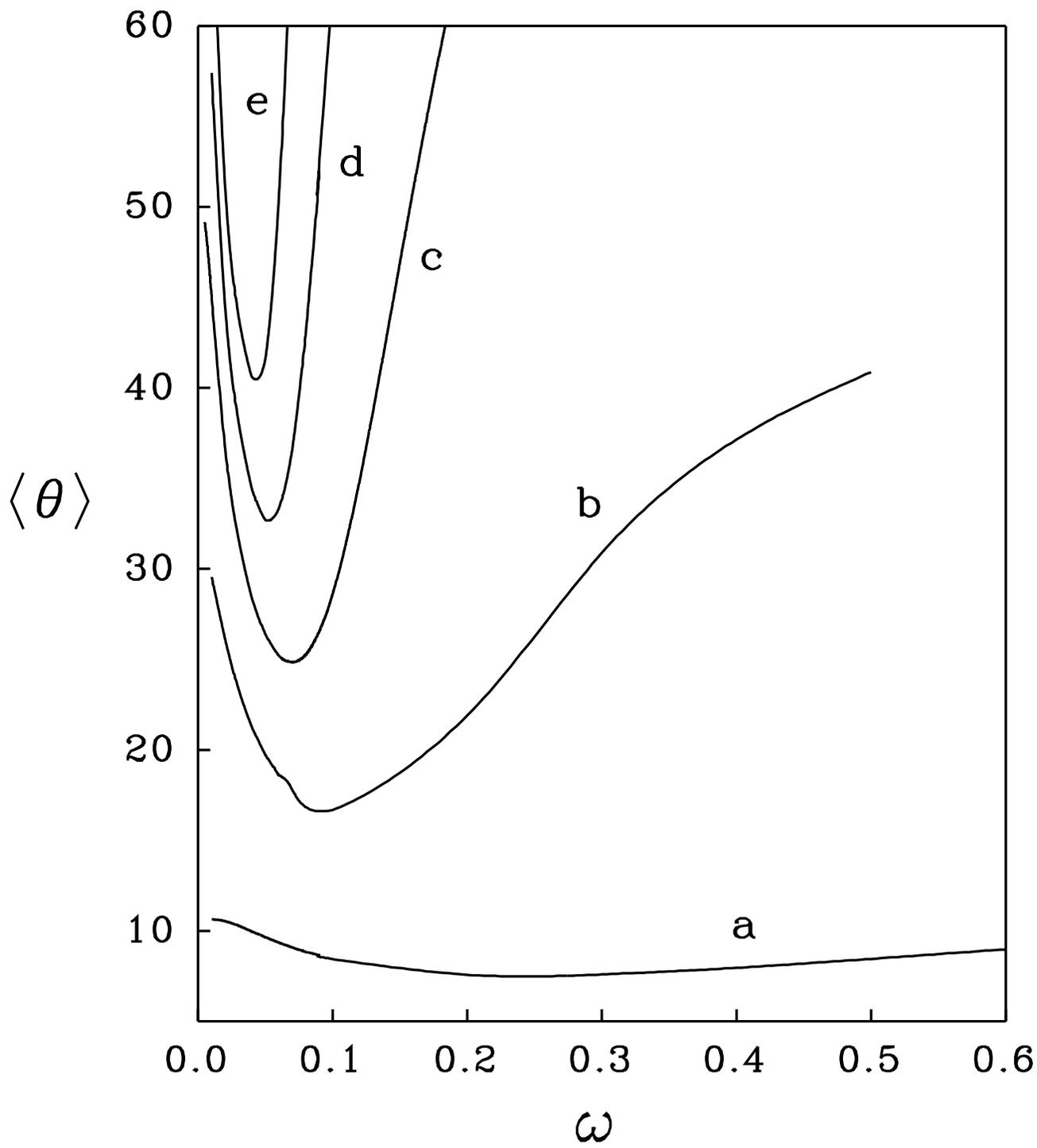
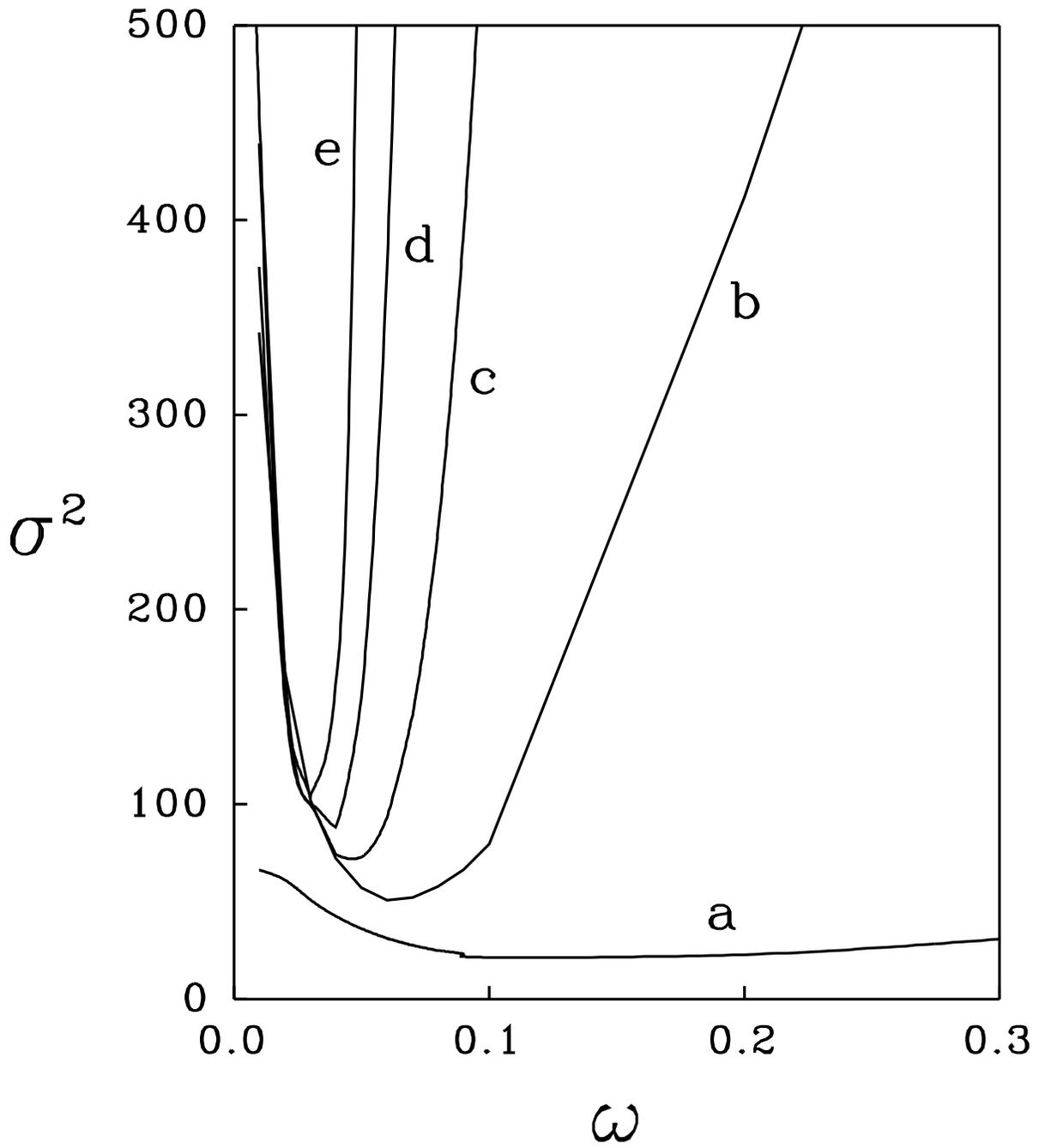
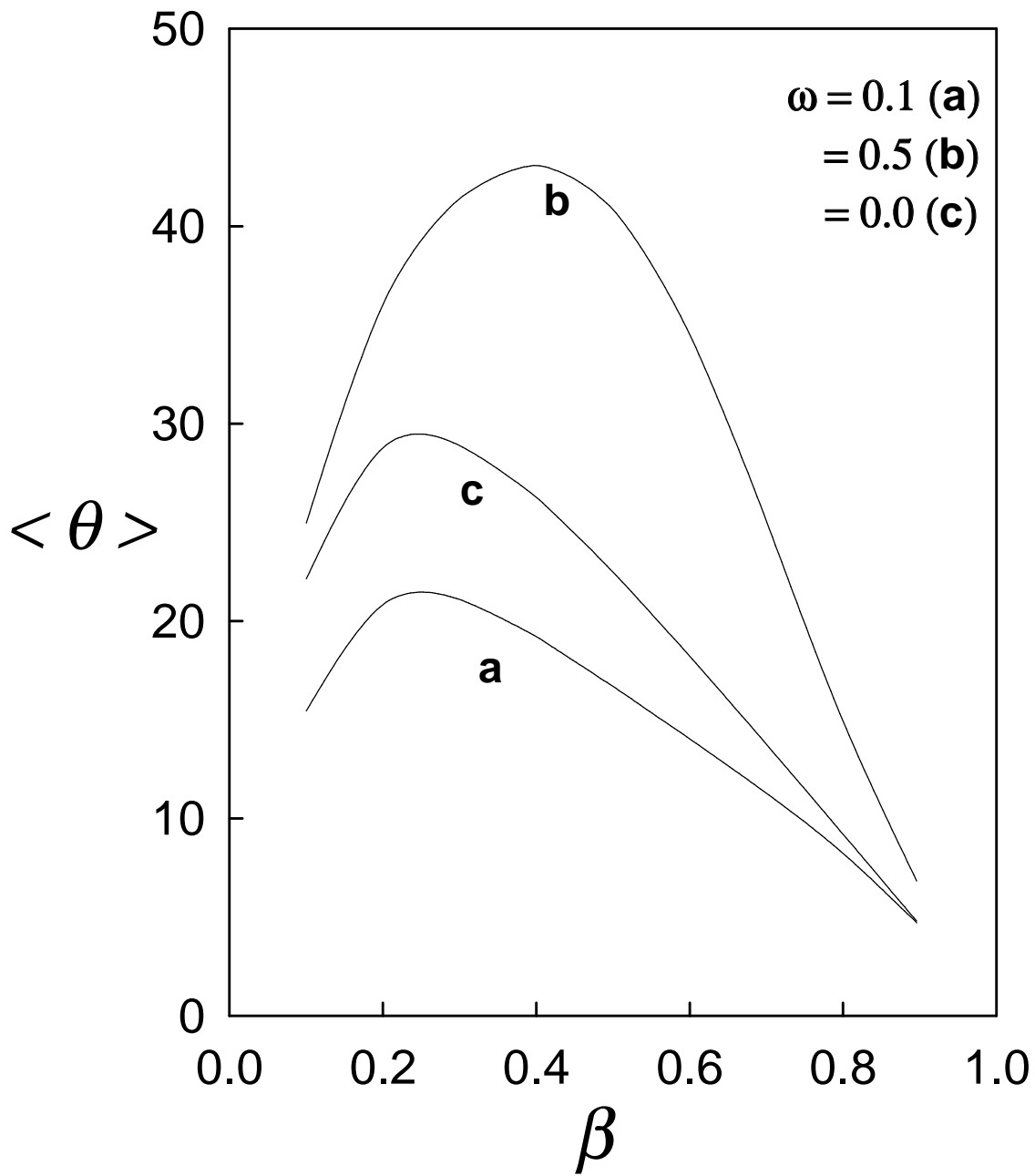
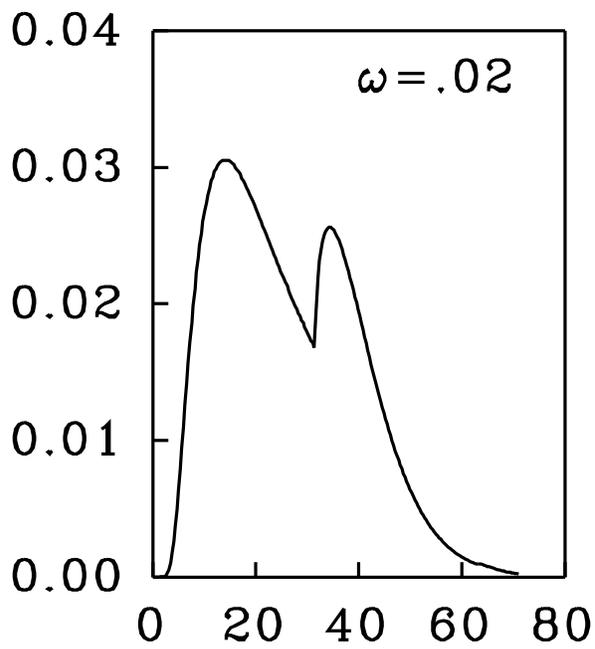
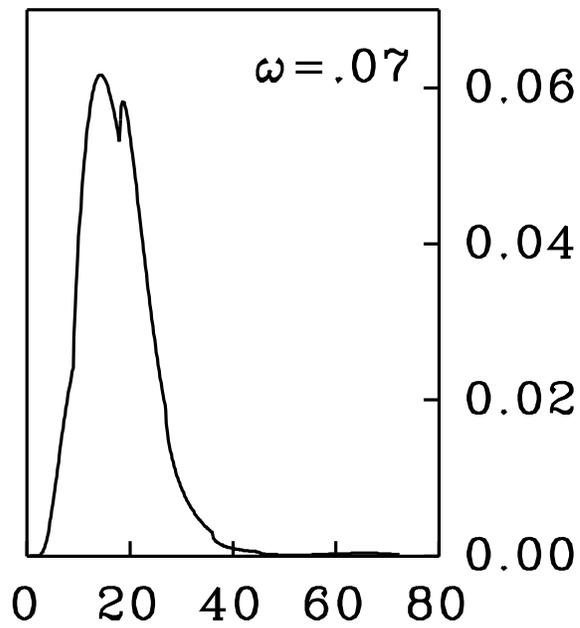
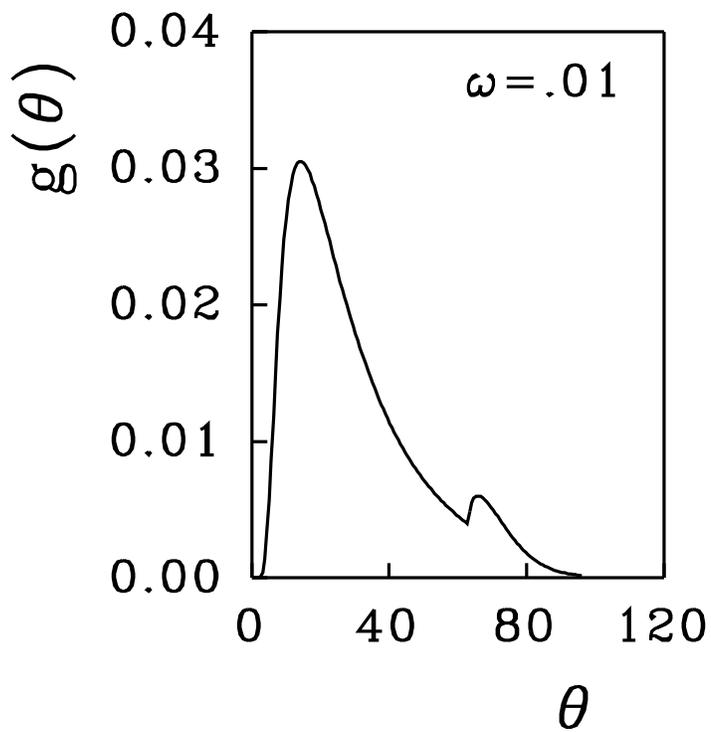
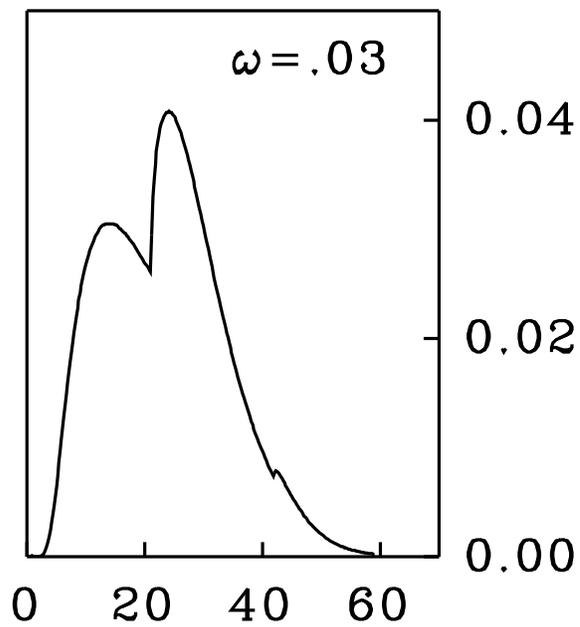
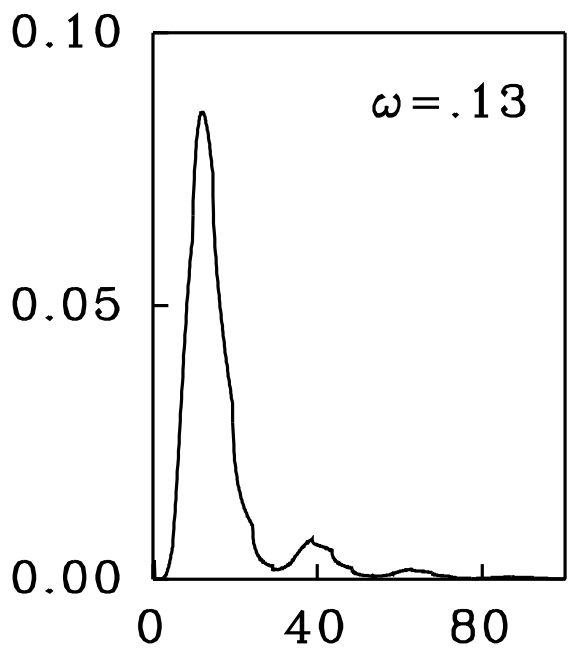
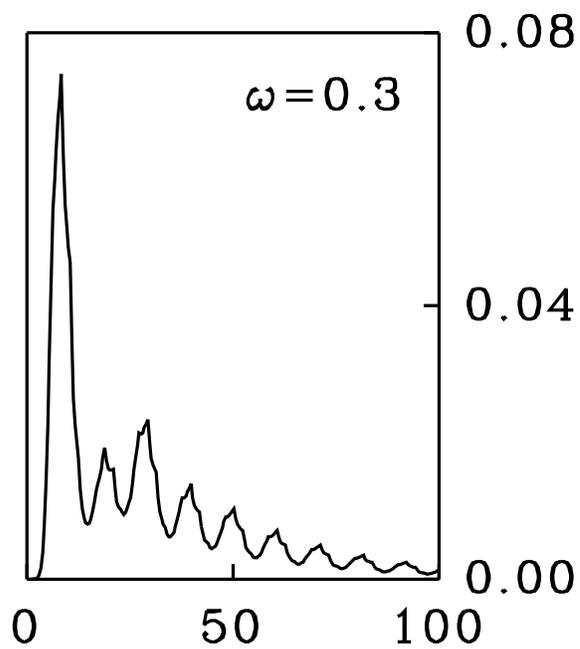
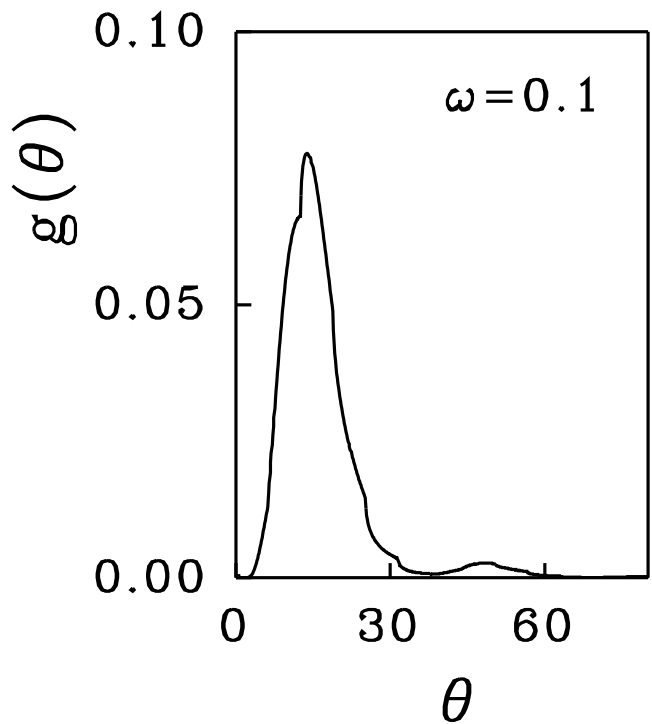
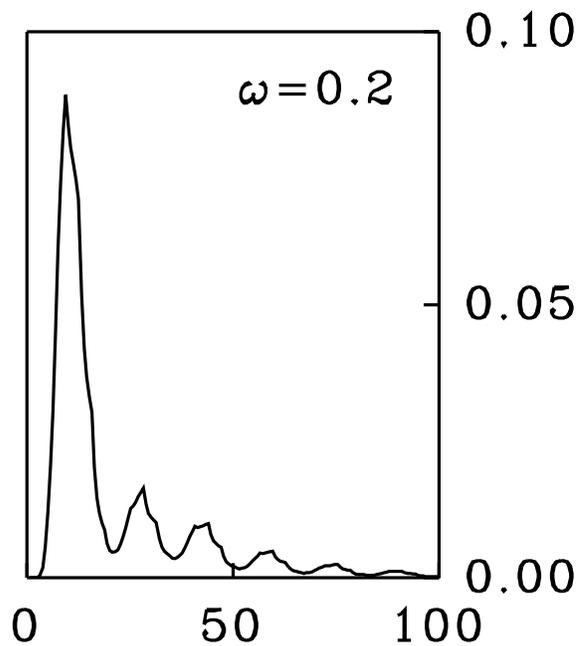
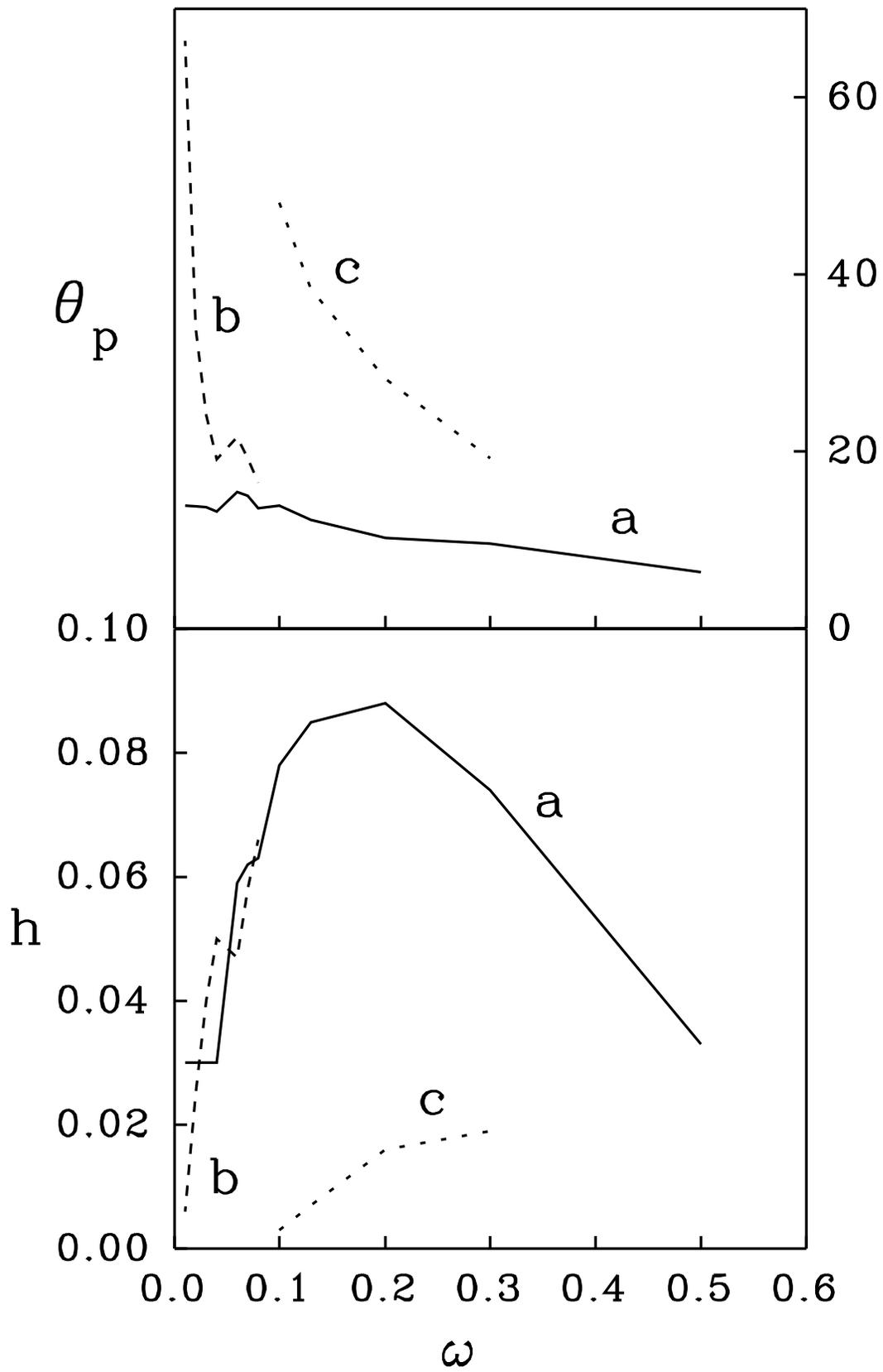
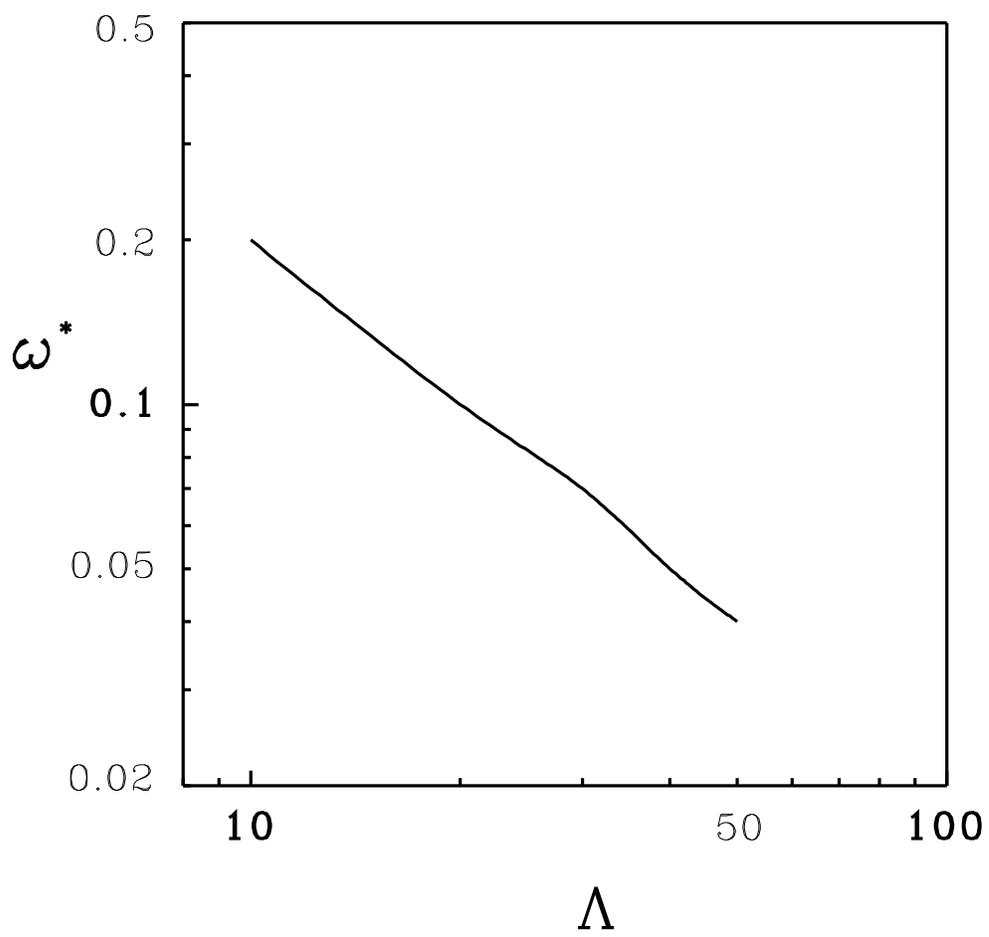
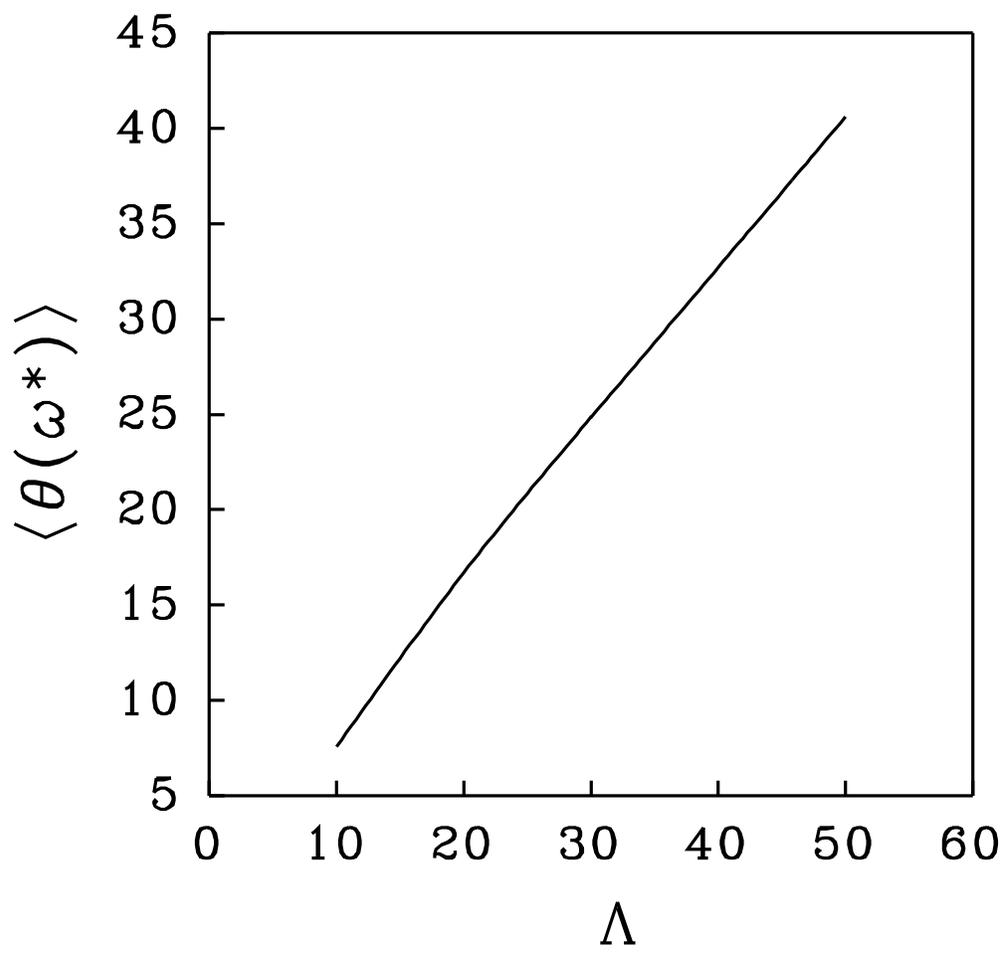
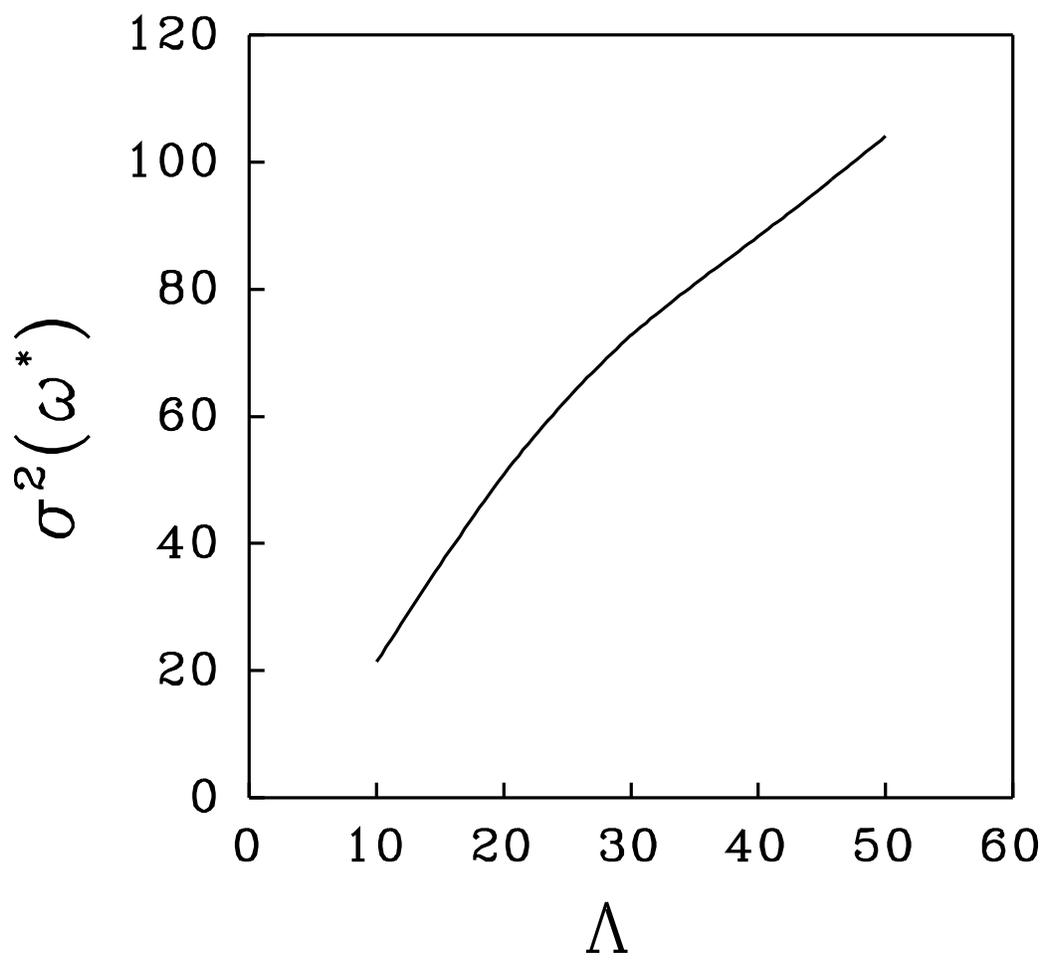
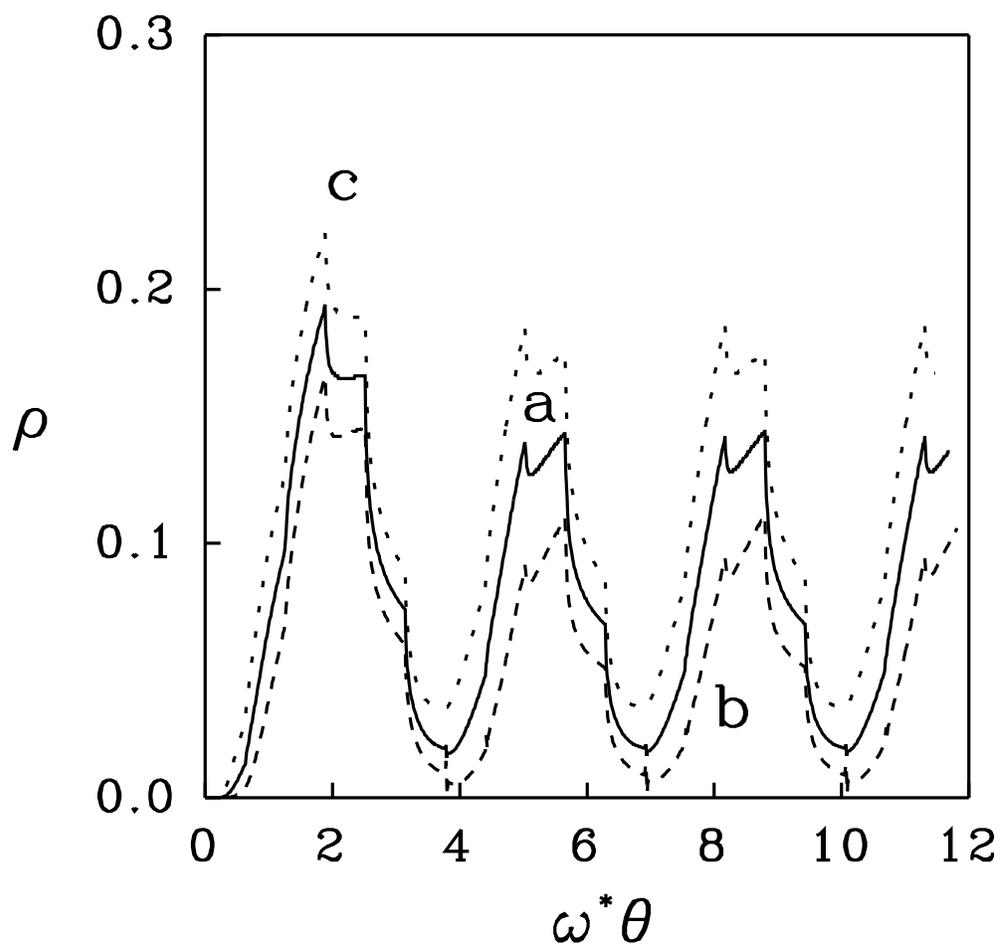
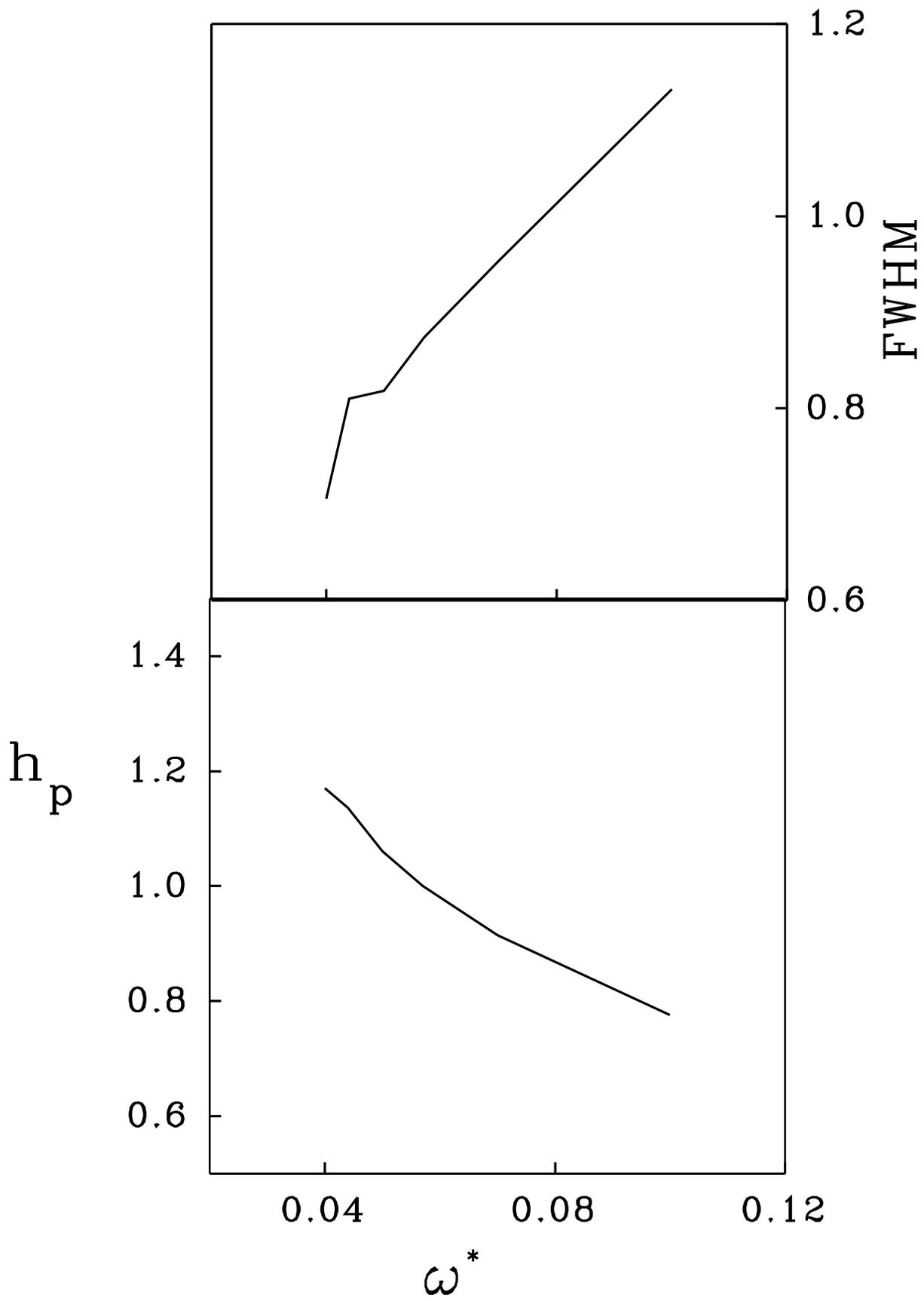
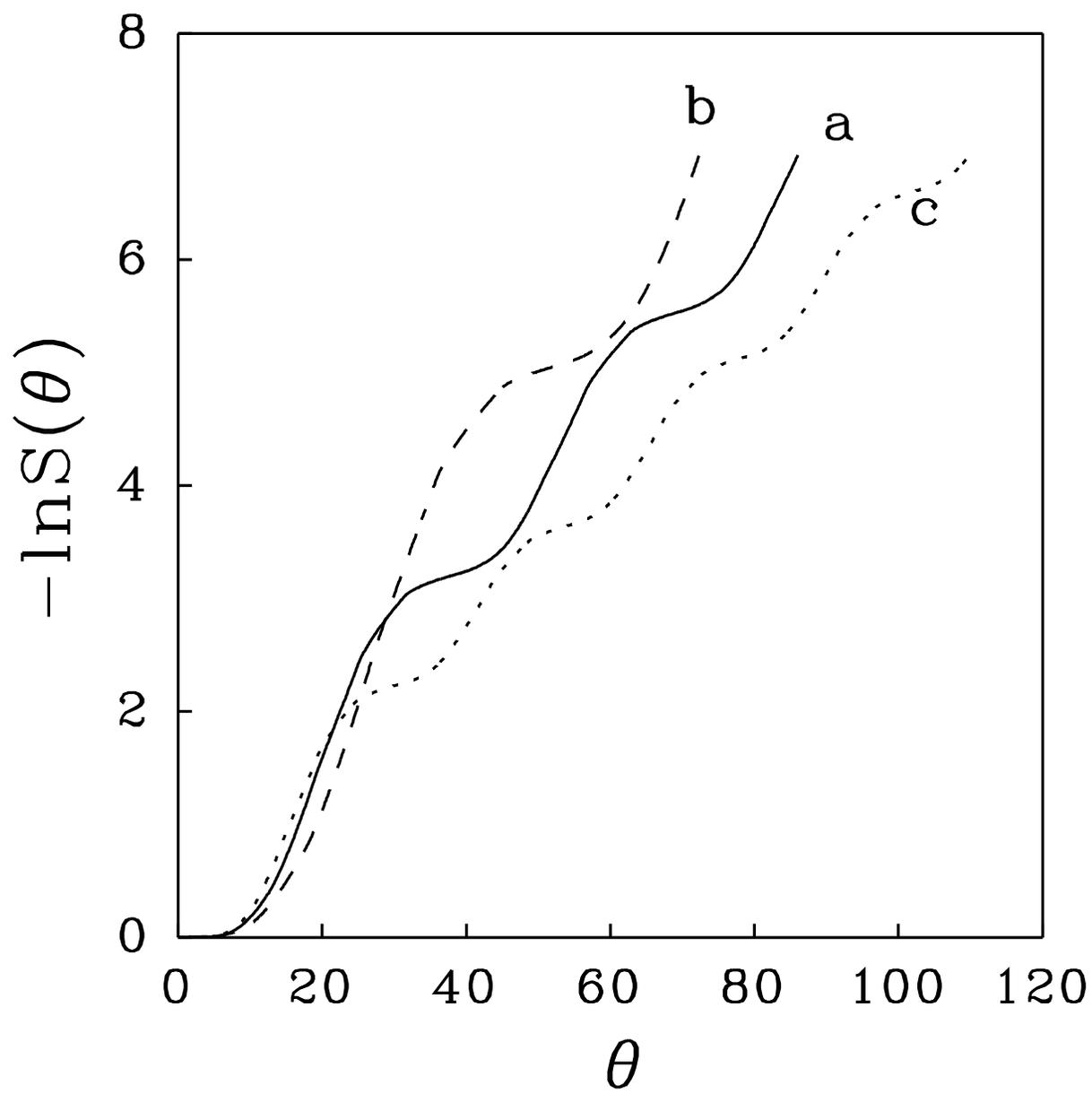
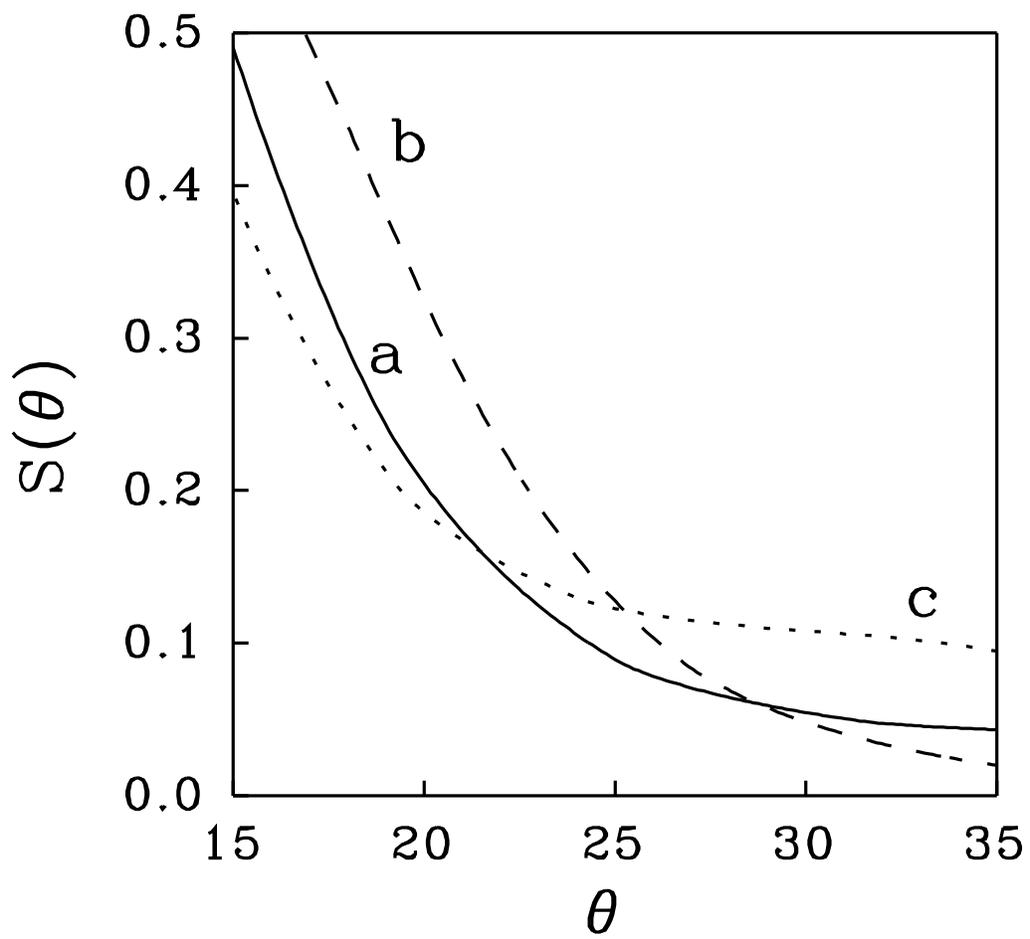
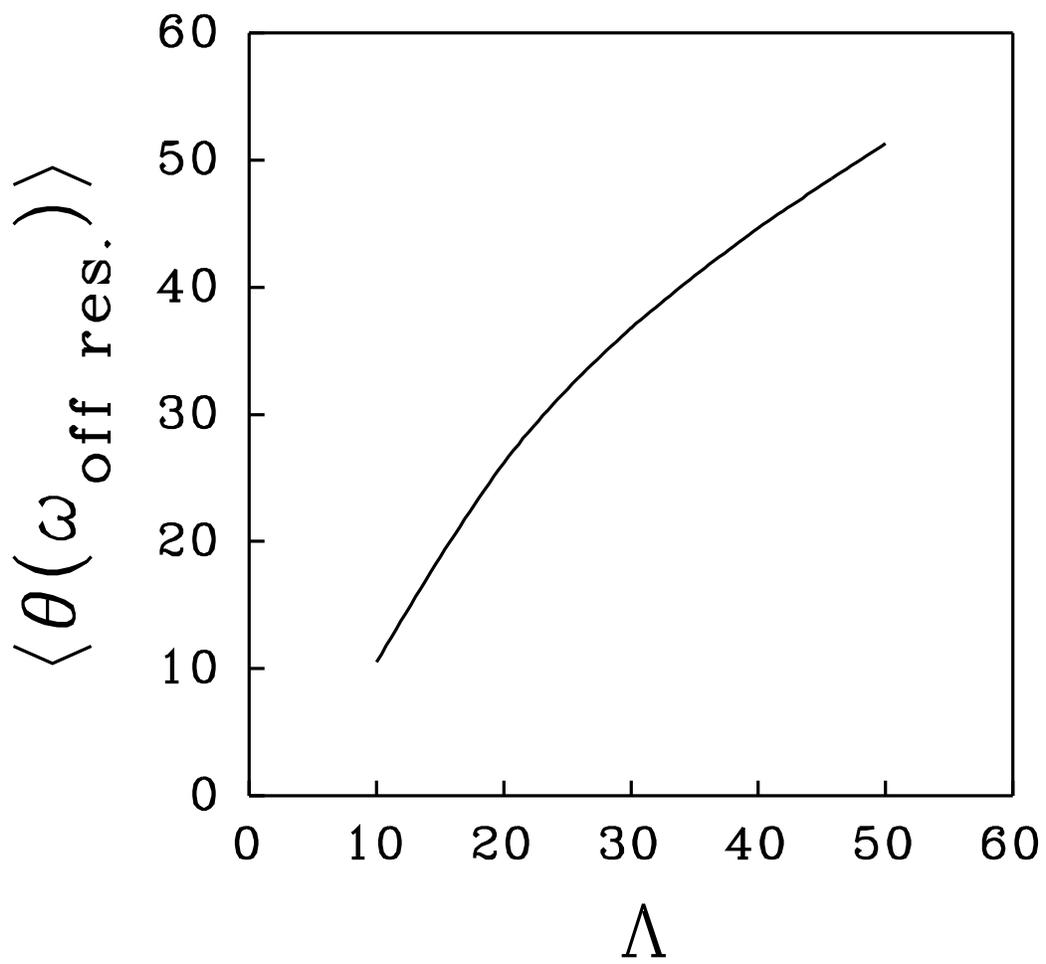
\end{document}